\documentclass[conference]{IEEEtran}

\usepackage{graphicx,subfigure}
\usepackage{amsmath,amssymb,amsthm}
\usepackage{pst-sigsys,pstricks-add}
\usepackage[noadjust]{cite}

\newtheorem{thm}{Theorem}
\newtheorem{lem}{Lemma}
\newtheorem{cor}{Corollary}

\theoremstyle{definition}

\newtheorem{example}{Example}

\newcommand{\Qvec}{\mathbf{Q}}
\newcommand{\nuvec}{\boldsymbol{\nu}}
\newcommand{\Lambdavec}{\boldsymbol{\Lambda}}
\newcommand{\Vvec}{\mathbf{V}}
\newcommand{\Uvec}{\mathbf{U}}
\newcommand{\Dvec}{\mathbf{D}}
\newcommand{\Rvec}{\mathbf{R}}

\newcommand{\ent}[1]{H(#1)}

\newcommand{\mutinf}[1]{I(#1)}

\newcommand{\kld}[2]{D(#1||#2)}

\newcommand{\loss}[2][\empty]{\ifthenelse{\equal{#1}{\empty}}{L(#2)}{L_{#1}(#2)}}
\newcommand{\lossrate}[2][\empty]{\ifthenelse{\equal{#1}{\empty}}{L(\mathbf{#2})}{L_{\mathbf{#1}}(\mathbf{#2})}}

\newcommand{\relLoss}[2][\empty]{\ifthenelse{\equal{#1}{\empty}}{l(#2)}{l_{#1}(#2)}}

\DeclareMathOperator*{\argmax}{arg\,max}

\newcommand{\dom}[1]{\mathcal{#1}}

\newcommand{\Prob}[1]{\mathrm{Pr}(#1)}

\newcommand{\diag}[1]{\mathrm{diag}(#1)}

\newcommand{\Pvec}{\mathbf{P}}
\newcommand{\muvec}{\boldsymbol{\mu}}
\newcommand{\wvec}{\mathbf{w}}
\newcommand{\Wvec}{\mathbf{W}}

\newcommand{\Amat}{\mathbf{A}}

\newcommand{\onevec}{\mathbf{1}}

\newcommand{\eye}{\mathbf{I}}

\newcommand{\muveca}{\muvec^{(1)}}
\newcommand{\muvecb}{\muvec^{(2)}}
\newcommand{\muveci}{\muvec^{(i)}}
\newcommand{\muvecj}{\muvec^{(j)}}
\newcommand{\gammavec}{\boldsymbol{\gamma}}

 \renewcommand{\mutinf}[1]{I\left(#1\right)}
 
\title{Hard Clusters Maximize Mutual Information}
\author{
\IEEEauthorblockN{Bernhard C. Geiger, Rana Ali Amjad}\\
\IEEEauthorblockA{Institute for Communications Engineering, Technical University of Munich, Germany\\
geiger@ieee.org, ranaali.amjad@tum.de}
}

\begin {document}
\maketitle

\begin{abstract}
 In this paper, we investigate mutual information as a cost function for clustering, and show in which cases hard, i.e., deterministic, clusters are optimal. Using convexity properties of mutual information, we show that certain formulations of the information bottleneck problem are solved by hard clusters. Similarly, hard clusters are optimal for the information-theoretic co-clustering problem that deals with simultaneous clustering of two dependent data sets. If both data sets have to be clustered using the same cluster assignment, hard clusters are \emph{not} optimal in general. We point at interesting and practically relevant special cases of this so-called pairwise clustering problem, for which we can either prove or have evidence that hard clusters are optimal. Our results thus show that one can relax the otherwise combinatorial hard clustering problem to a real-valued optimization problem with the same global optimum.
\end{abstract}


\section{Introduction}\label{sec:intro}
Information-theoretic clustering, i.e., clustering employing information-theoretic cost functions, has received a lot of attention in the last decade. Information-theoretic cost functions capture more than just linear correlation, allowing to identify nonlinear dependencies between data points and to obtain clusters with a more complicated structure than, e.g., k-means is capable of. Thus, the proposed information-theoretic methods were shown to perform as good as or even better than competing spectral, centroid-based, hierarchical, or model-based clustering methods.

A particularly successful example is the information bottleneck method~\cite{Tishby_InformationBottleneck}. Motivated by rate-distortion theory with an information-theoretic distortion function, it tries to cluster data into as few clusters as possible, while preserving as much information as possible about a ``relevant random variable''. Taken as a pre-processing step for classification, the data could be a random feature vector, that should be compressed while staying informative about a class random variable. Both compression and information-preservation are quantified by mutual information.

The capability of mutual information to capture nonlinear statistical dependencies subsequently made it an attractive cost function for
tackling various types of clustering problems: Information-theoretic clustering algorithms similar to the information bottleneck method have been proposed~\cite{Strouse_DIB}, and there are algorithms for simultaneous clustering of two dependent data sets~\cite{Dhillon_ITClustering}, for obtaining clusters that are in some sense orthogonal to a predefined clustering~\cite{Gondek_Conditional}, and for clustering data sets based on the pairwise similarities between data points~\cite{Alush_PairwiseClustering,Tishby_MarkovRelax}. 

Despite its success, we believe that mutual information as a cost function for clustering, and the proposed algorithms to optimize it, are not well understood. In this paper, we fill this gap by analyzing several settings of information-theoretic clustering. 
In Section~\ref{sec:notation} we build the bridge between a data set and its probabilistic description, and introduce properties of mutual information relevant for this work. We then turn to the problem of simultaneous clustering two dependent data sets, dubbed co-clustering (Section~\ref{sec:cocluster}), and to a specific formulation of the clustering problem related to the information bottleneck problem (Section~\ref{sec:cluster}). We show that information-theoretic cost functions for these problems are maximized by deterministic, i.e., hard clusters: The optimal solution of the \emph{real-valued} soft clustering problem coincides with the solution of the \emph{combinatorial} hard clustering problem. As a consequence, we argue that the heuristic methods proposed in the literature to solve the latter can be complemented by methods targeted at solving the former. Future work shall mainly deal with developing information-theoretic clustering algorithms based on this observation.

One can restrict the information-theoretic co-clustering problem such that the cluster assignments for the two data sets are identical. This approach is used for clustering data sets characterized by pairwise (dis-)similarities, and is hence called pairwise clustering. We show in Section~\ref{sec:pairwisecluster} that for this scenario, hard clusters need not be optimal in general. In contrast, Section~\ref{sec:whenhard} discusses several practically relevant special cases for which we can either prove, or have convincing numerical evidence, that the optimal pairwise clusters are hard. 


\section{Preliminaries}\label{sec:notation}
\subsection{The Probabilistic Approach to Clustering}
In a clustering problem, the aim is to find a (possibly stochastic) mapping $f^\bullet$ of a set of $N$ elements to a set of $M<N$ elements based on available data $\mathcal{D}$ that is optimal w.r.t.\ some cost function $\mathbb{C}(\mathcal{D},f)$.

In a co-clustering problem, the aim is the simultaneous clustering of two dependent sets. We look, based on the available data $\mathcal{D}$, for two mappings $f_1^\bullet$ and $f_2^\bullet$, each clustering 
one of the two sets optimally w.r.t.\  some cost function $\mathbb{C}(\mathcal{D},f_1,f_2)$.

The probabilistic approach to (co-)clustering first transforms the data $\mathcal{D}$ into a probabilistic description $\mathcal{P}$ before clustering. I.e., it consists of the following two steps:
\begin{itemize}
 \item Define a probabilistic model $\mathcal{P}$ for the data $\mathcal{D}$. 
 \item Optimize some cost function $\mathbb{C}(\mathcal{P},f)$ (or $\mathbb{C}(\mathcal{P},f_1,f_2)$) to find the optimal (co-)clustering $f$ ($f_1$, $f_2$).
\end{itemize}
The method to define the model $\mathcal{P}$ can depend on the type of available data $\mathcal{D}$, the specific application of (co-)clustering in mind, and the cost function being used. This two step approach
has the benefit that we can adapt the transformation of the data to $\mathcal{P}$ according to the specific application while keeping the cost function fixed. Hence we can develop efficient algorithms 
to solve the optimization problem for a fixed cost function independently of the underlying application. 

A popular example for the probabilistic approach is clustering based on Gaussian mixture models (GMMs): In the first step, the expectation-maximization algorithm is used to model the data by a GMM which is used in the second step to obtain the mapping $f$.

\subsection{From Data to Probabilities}\label{subsec:probabilistic} 
Since this work considers mutual information as a cost function, we restrict our attention to the case where the probabilistic model $\mathcal{P}$ contains the joint distribution between two random variables (RVs). Specifically, if $X_1$ and $X_2$ are RVs with alphabets $\dom{X}_1=\{1,2,\dots,N_1\}$ and $\dom{X}_2=\{1,2,\dots,N_2\}$, their joint distribution is given by an $N_1\times N_2$ matrix $\Pvec=\{P_{kl}\}$, where $P_{kl}:=\Prob{X_1=k,X_2=l}$. While in the rest of the paper we assume that $\Pvec$ is given, we mention two examples that illustrate how $\Pvec$ can be obtained from data $\mathcal{D}$.

\begin{example}\label{exp:pairwise}
 Let the data $\mathcal{D}$ consist of $N$ points $\textbf{a}_i \in \mathbb{R}^k$, and let $d_{ij}$ be the Euclidean distance between the $i$-th and the $j$-th point. Using these pairwise distances, we write
 \begin{align}
     \mathbf{P} &= \frac{1}{\sum\limits_{i,j} e^{-\sigma d_{ij}}}
     \begin{bmatrix} 
     e^{-\sigma d_{11}} &\cdots & e^{-\sigma d_{1N}} \\ 
     \vdots & \ddots & \vdots \\ 
     e^{-\sigma d_{N1}} & \cdots & e^{-\sigma d_{NN}} 
     \end{bmatrix}
 \end{align}
 where $\sigma>0$ is a scaling parameter.
\end{example}

\begin{example}\label{exp:co}
 Let the available data $\mathcal{D}$ consist of a set of $N_1$ books, containing a total of $N_2$ words. From $\mathcal{D}$ we can build a relationship matrix $\mathbf{D}$, where $D_{ij}$ is the number of times the $j$-th word occurs in the $i$-th book. We obtain a joint probability 
 matrix $\mathbf{P}$ by normalizing $\mathbf{D}$,
 \begin{align}
  \mathbf{P} &= \frac{1}{\overline{\onevec}\textbf{D}\onevec} \textbf{D}
 \end{align}
 where $\onevec$ represents a column vector of ones and $\overline{\onevec}$ represents its transpose.
\end{example}

\subsection{Soft vs. Hard Clustering}
Soft (or stochastic) clustering means that the mapping $f$ (or $f_1$ and $f_2$) assigns, to each data point $i$ to be clustered, a probability vector $\overline{\mathbf{w}_i} = \begin{bmatrix} W_{i1} & \cdots & W_{iM}\end{bmatrix}$, where 
$W_{ik}$ represents the probability of data point $i$ being a part of the cluster $k$. Hard (or deterministic) clustering puts each data point $i$ into exactly one cluster $k$, i.e., for each $i$ there is one $k$ such that $W_{ik} =1$, while for all other $j\neq k$, $W_{ij} =0$. Hence soft clustering can be viewed as a stochastic channel from data points to the clusters, and hard clustering is a special case of it.

Stacking the vectors $\overline{\mathbf{w}_i}$ in an $N\times M$ matrix $\Wvec$, we get a matrix describing the conditional distribution of a cluster RV $Y$ given a data point RV $X$. Since $\Wvec\onevec=\onevec$, $\Wvec$ is an element of the space $\mathcal{M}_{N\times M}$ of $N\times M$ stochastic matrices. \emph{Deterministic} matrices $\Vvec\in\mathcal{M}_{N\times M}$, corresponding to hard clusters, are special cases of stochastic matrices and satisfy $V_{kl}\in\{0,1\}$ for all $k,l$. Every stochastic matrix is the convex combination of finitely many deterministic matrices. 
\begin{lem}[{\cite[Thm.~1]{Davis_ConvexCombination}}]\label{lem:convcomb}
Every stochastic matrix $\Wvec\in\mathcal{M}_{N\times M}$ can be written as the convex combination of at most $N(M-1)+1$ deterministic matrices, i.e.,
\begin{equation}
 \Wvec=\sum_{i=1}^{N(M-1)+1} \lambda_i \Vvec^{(i)}
\end{equation}
where $\lambda_i\ge 0$ and $\sum_{i=1}^{N(M-1)+1} \lambda_i=1$.
\end{lem}
In other words, $\mathcal{M}_{N\times M}$ is a convex polytope with deterministic matrices as its vertices. This fact is closely related to the Birkhoff-von~Neumann theorem, a constructive proof of which appears in~\cite[Thm.~2.1.6]{Bapat_NonnegativeMatrices}.

\subsection{Mutual Information as a Cost Function}\label{subsec:}
As discussed in Section~\ref{sec:intro}, mutual information has become an important cost function for clustering because of its capability to capture non-linear dependencies. We now introduce some of its properties.

The joint and marginal distributions of $X_1$ and $X_2$ are given as $\Pvec$, $\muveca=\Pvec\onevec$, and $\overline\muvecb=\overline\onevec\Pvec$. Let $\diag{\muveci}$ be a diagonal matrix with the entries of $\muveci$ on the main diagonal. With this notation, we write $\mutinf{\Pvec}:=\mutinf{X_1;X_2}$ for the mutual information between $X_1$ and $X_2$, $\ent{\muveci}:=\mutinf{\diag{\muveci}}=\ent{X_i}$ for the entropy of $X_i$, and $\kld{\muvec}{\muvec'}$ for the Kullback-Leibler divergence between the two distributions $\muvec$ and $\muvec'$~\cite[Ch.~2]{Cover_Information2}. 

\begin{lem}[{\cite[Thm.~2.7.2]{Cover_Information2}}]\label{lem:kldconvex}
 The Kullback-Leibler divergence $\kld{\muvec}{\muvec'}$ between two distributions $\muvec$ and $\muvec'$ is convex in the pair $(\muvec,\muvec')$, i.e., 
 \begin{multline}
  \kld{\lambda\muveca+(1-\lambda)\muvecb}{\lambda\muvec^{'(1)}+(1-\lambda)\muvec^{'(2)}}\\
  \le \lambda\kld{\muveca}{\muvec^{'(1)}} + (1-\lambda)\kld{\muvecb}{\muvec^{'(2)}}.
 \end{multline}
\end{lem}

\begin{cor}[{\cite[Thm.~2.7.4]{Cover_Information2}}]\label{cor:miconvex}
 The mutual information $\mutinf{\diag{\muvec}\Wvec}$ is convex in $\Wvec$ for fixed $\muvec$ and concave in $\muvec$ for fixed $\Wvec$. Moreover,  $\mutinf{\diag{\overline{\muvec}\Wvec}}=\ent{\overline{\muvec}\Wvec}$ is concave in $\Wvec$ for fixed $\muvec$.
\end{cor}

In what follows, we will show scenarios in which mutual information as a cost function is optimized by hard clusters. Some of our derivations are based on fact that a convex function over $\mathcal{M}_{N\times M}$ attains its maximal value at a deterministic matrix $\Vvec$. This is a direct consequence of Lemma~\ref{lem:convcomb} and the following result:
\begin{lem}[{\cite[Ch.~4]{Benson_Concave}}]\label{lem:convmax}
 A convex function over a convex polytope attains its maximal value at one of the polytope's vertices.
\end{lem}

\section{Hard Co-Clusters are Optimal}\label{sec:cocluster}
\begin{figure}[t]
 \centering
    \begin{pspicture}[showgrid=false](-4,0.25)(4,1.5)
      \psset{style=RoundCorners}
      \pssignal(-4,1){y1}{$Y_1$}
      \pssignal(-1,1){x1}{$X_1$}
      \pssignal(1,1){x2}{$X_2$}
      \pssignal(4,1){y2}{$Y_2$}
      \ncline{x1}{x2}
      \psfblock[framesize=1 0.75](2.5,1){w2}{$\Wvec^{(2)}$}
      \psfblock[framesize=1 0.75](-2.5,1){w1}{$\Wvec^{(1)}$}
      \nclist[style=Arrow]{ncline}{x2,w2,y2} 
      \nclist[style=Arrow]{ncline}{x1,w1,y1} 
      \ncarc[style=Dash,arcangleA=-90,arcangleB=-90,linecolor=red]{x1}{x2}
      \rput*(0,0.1){\textcolor{red}{$\Pvec$}}
    \end{pspicture}
\caption{Deterministic clusters $\Wvec^{(1)}$ and $\Wvec^{(2)}$ maximize $\mutinf{Y_1;Y_2}$ in the co-clustering problem (cf.~Theorem~\ref{thm:coclustering}).}
\label{fig:coclusters}
\end{figure}
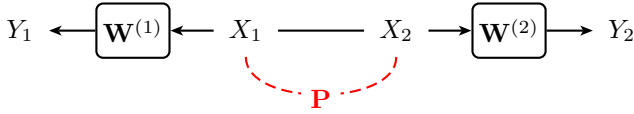

We depict the information-theoretic formulation of co-clustering in Fig.~\ref{fig:coclusters}. Our aim is to find mappings $\Wvec^{(1)}$ and $\Wvec^{(2)}$ between the RVs $X_1$ and $Y_1$, and between $X_2$ and $Y_2$, respectively. In the framework of Example~\ref{exp:co}, $X_1$ is a RV over books, $X_2$ a RV over words, and the aim of information-theoretic co-clustering is to maximize the mutual information between the clusters $Y_1$ of books and the clusters $Y_2$ of words. 

Algorithms for information-theoretic co-clustering date back to Dhillon et al.~\cite{Dhillon_ITClustering}. The authors restricted themselves to hard clustering and presented a suboptimal, sequential algorithm. We will show in this section that the restriction to hard clusters comes without loss of generality, and that hence the problem can be attacked with a larger set of tools.

Formulating the problem mathematically, we are interested in finding a maximizer $(\Wvec^{(1)\bullet},\Wvec^{(2)\bullet})$ of
\begin{equation}\label{eq:problem_coclustering}
 \max_{\substack{\Wvec^{(1)}\in\mathcal{M}_{N_1\times M_1}\\\Wvec^{(2)}\in\mathcal{M}_{N_2\times M_2}}} \mutinf{\overline{\Wvec^{(1)}}\Pvec\Wvec^{(2)}}.
\end{equation}

\begin{thm}\label{thm:coclustering}
 At least one solution $(\Wvec^{(1)\bullet},\Wvec^{(2)\bullet})$ to problem~\eqref{eq:problem_coclustering} consists of two deterministic matrices.
\end{thm}
\begin{IEEEproof}
 The proof follows again by applying Lemma~\ref{lem:convcomb}. Assuming that there is no deterministic solution, we decompose the stochastic solutions $\Wvec^{(j)\bullet}$, $j=1,2$, as
 \begin{equation}
  \Wvec^{(j)\bullet} = \sum_{i=1}^{r_j} \lambda^{(j)}_i \Vvec^{(j,i)}
 \end{equation}
 where $r_j\le N_j(M_j-1)+1$. Careful calculations show that $\Qvec:=\overline{\Wvec^{(1)\bullet}}\Pvec\Wvec^{(2)\bullet}$ can thus be written as
\begin{equation}
 \Qvec = \sum_{i=1}^{r_1}\sum_{k=1}^{r_2} \lambda^{(1)}_i \lambda^{(2)}_k \underbrace{\overline{\Vvec^{(1,i)}}\Pvec\Vvec^{(2,k)}}_{=:\Qvec^{(i,k)}}.
\end{equation}
While $\Qvec$ denotes the joint distribution of $Y_1$ and $Y_2$, we let $\tilde\Qvec:=\nuvec^{(1)}\overline{\nuvec^{(2)}}$ denote the product of their marginal distributions. But with
\begin{equation}
 \overline{\nuvec^{(j)}}=\overline{\muvecj}\Wvec^{(j)\bullet}=\sum_{i=1}^{r_j} \lambda^{(j)}_i \overline{\muvecj}\Vvec^{(j,i)}
\end{equation}
we get
\begin{equation}
 \tilde\Qvec=\nuvec^{(1)}\overline{\nuvec^{(2)}} = \sum_{i=1}^{r_1}\sum_{k=1}^{r_2} \lambda^{(1)}_i \lambda^{(2)}_k \underbrace{\overline{\Vvec^{(1,i)}}\muveca \overline{\muvecb}\Vvec^{(2,k)}}_{=:\tilde\Qvec^{(i,k)}}.
\end{equation}
We can now write the mutual information between $Y_1$ and $Y_2$ as the Kullback-Leibler divergence between their joint distribution and the product of their marginal distributions and get
\begin{align}
 \mutinf{\Qvec} &= \kld{\Qvec}{\tilde{\Qvec}}\\
 &\stackrel{(a)}{\le} \sum_{i=1}^{r_1}\sum_{k=1}^{r_2} \lambda^{(1)}_i \lambda^{(2)}_k \kld{\Qvec^{(i,k)}}{\tilde{\Qvec}^{(i,k)}}\\
 &= \sum_{i=1}^{r_1}\sum_{k=1}^{r_2} \lambda^{(1)}_i \lambda^{(2)}_k \mutinf{\overline{\Vvec^{(1,i)}}\Pvec\Vvec^{(2,k)}}
\end{align}
where $(a)$ is due to Lemma~\ref{lem:kldconvex}. Since the right-hand side of the last equation is a convex combination of non-negative terms, at least one of them is equal to or greater than $\mutinf{\Qvec}$. Hence, at least one deterministic pair $(\Vvec^{(1,i)},\Vvec^{(2,k)})$ in the decomposition is a solution to the problem in~\eqref{eq:problem_coclustering}. 
\end{IEEEproof}

This results allows to relax the hard co-clustering problem, which is NP-hard, to a continuous-valued problem. Although the problem~\eqref{eq:problem_coclustering} is not convex, there is hope that heuristic algorithms for~\eqref{eq:problem_coclustering} perform better than the combinatorial algorithm proposed in~\cite{Dhillon_ITClustering} for hard co-clustering. Future work shall evaluate this claim.

This result also shows that if one aims to find soft co-clusters, the mutual information $\mutinf{Y_1;Y_2}$ is not an appropriate cost function.

\section{Information Bottleneck-Based Clustering leads to Hard Clusters}\label{sec:cluster}
\begin{figure}[t]
 \centering
    \begin{pspicture}[showgrid=false](-2,0.25)(3,1.5)
      \psset{style=RoundCorners}
      \pssignal(-2,1){x1}{$X_1$}
      \pssignal(-0,1){x2}{$X_2$}
      \pssignal(3,1){y2}{$Y_2$}
      \ncline{x1}{x2}
      \psfblock[framesize=1 0.75](1.5,1){w2}{$\Wvec$}
      \nclist[style=Arrow]{ncline}{x2,w2,y2} 
      \ncarc[style=Dash,arcangleA=-90,arcangleB=-90,linecolor=red]{x1}{x2}
      \rput*(-1,0.1){\textcolor{red}{$\Pvec$}}
    \end{pspicture}
\caption{A deterministic cluster $\Wvec$ maximizes $\mutinf{X_1;Y_2}$ in the clustering problem (cf.~Theorem~\ref{thm:clustering}). Clustering to maximize $\mutinf{X_2;Y_2}$ is subsumed by setting $\Pvec=\diag{\muvecb}$.}
\label{fig:clusters}
\end{figure}
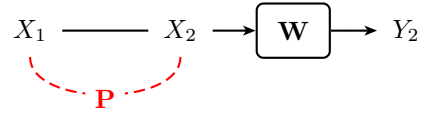

We are now interested in the problem depicted in Fig.~\ref{fig:clusters}. Our aim is to find a mapping $\Wvec$ between the RVs $X_2$ and $Y_2$. In the framework of Example~\ref{exp:co}, $X_1$ is a RV over books and $X_2$ is a RV over words, but now we are only interested in clustering words into clusters $Y_2$ such that the mutual information between books $X_1$ and word clusters $Y_2$ is maximized. Formulating the problem mathematically, we are interested in finding a maximizer $\Wvec^\bullet$ of
\begin{equation}
 \max_{\Wvec\in\mathcal{M}_{N_2\times M_2}} \mutinf{\Pvec\Wvec}.\label{eq:problem_clustering}
\end{equation}

Note the similarity between this problem and the information bottleneck method~\cite{Tishby_InformationBottleneck}, which tries to solve
\begin{equation}
 \max_{\Wvec\in\mathcal{M}_{N_2\times \cdot}}  \mutinf{\Pvec\Wvec} - \beta \mutinf{\diag{\muvecb}\Wvec}.
\end{equation}
In other words, the information bottleneck method does not explicitly limit the number of clusters, but rather restricts the mutual information shared between $X_2$ and the clustered RV $Y_2$. The solution to this problem is obtained by the following implicit equation:
 \begin{align}\label{eq:IBeq}
 W_{ij} &\propto (\overline\onevec\Pvec\Wvec)_j e^{-\beta \kld{\wvec_i^{(1|2)}}{(\Pvec\Wvec)_{\cdot j}/(\overline\onevec\Pvec\Wvec)_j}} 
\end{align}
where $\overline{\wvec_i^{(1|2)}}$ is the $i$-th row of $\Wvec^{(1|2)}=\diag{\muvecb}^{-1}\overline{\Pvec}$.
While in general the solution is stochastic, these equations reveal that in the limit of $\beta\to\infty$ an optimal solution is deterministic. We now show, based on more elementary results, that a deterministic solution is optimal for problem~\eqref{eq:problem_clustering}:

\begin{thm}\label{thm:clustering}
 At least one solution $\Wvec^\bullet$ to problem~\eqref{eq:problem_clustering} is a deterministic matrix.
\end{thm}

\begin{IEEEproof}
We decompose $\Pvec$ as $\Pvec=\diag{\muveca}\Wvec^{(2|1)}$ to get
 \begin{equation}
  \mutinf{\Pvec\Wvec} = \mutinf{\diag{\muveca}\Wvec^{(2|1)}\Wvec}.
 \end{equation}
By Lemma~\ref{lem:convcomb}, $\Wvec$ can be written as a convex combination of deterministic matrices, and with Corollary~\ref{cor:miconvex} we obtain
\begin{align}
 \mutinf{\Pvec\Wvec} &= \mutinf{\diag{\muveca}\Wvec^{(2|1)}\sum_{i=1}^r \left(\lambda_i \Vvec^{(i)}\right)}\\
 &\le \sum_{i=1}^r \lambda_i \mutinf{\diag{\muveca}\Wvec^{(2|1)}\Vvec^{(i)}}
\end{align}
for $r\le N_2(M_2-1)+1$. This shows that $\mutinf{\Pvec\Wvec}$ is convex in $\Wvec$. Lemma~\ref{lem:convmax} completes the proof.
\end{IEEEproof}

A simple corollary of this result applies if one wants to restrict the alphabet of a RV while preserving most of its information, i.e., if one is looking for a $\Wvec\in\mathcal{M}_{N_2\times M_2}$ that maximizes $\mutinf{X_2;Y_2}$. Setting $\Pvec=\diag{\muvecb}$ in \eqref{eq:problem_clustering} reduces it to this problem. Hence, by Theorem~\ref{thm:clustering}, we know that hard clustering is optimal.

Since the sum of two convex functions is convex, this proof can be used to show that the problem
\begin{equation}
 \max_{\Wvec\in\mathcal{M}_{N_2\times \cdot}}  \mutinf{\Pvec\Wvec} - \beta \ent{\overline\muvecb\Wvec}
\end{equation}
is solved by at least one deterministic matrix $\Wvec^\bullet$. The authors of~\cite{Strouse_DIB} investigated this problem, provided a slightly more difficult proof for its solution being deterministic, and dubbed the problem thus the \emph{deterministic information bottleneck} problem. As in the original information bottleneck problem, also here an implicit equation is obtained as the solution, and the authors formulated an iterative procedure to find an optimum.

Theorem~\ref{thm:clustering} is not surprising, as one can view~\eqref{eq:problem_clustering} as a special case of~\eqref{eq:problem_coclustering} where we fix $\Wvec^{(1)}$ to be identity matrix. The particular appeal of the results in this section, however, is not only that the real-valued optimization problem is equivalent to a combinatorial one, but that, unlike~\eqref{eq:problem_coclustering}, the cost functions studied in this section are convex. Hence, although the problem is NP-hard, we can apply tools from non-convex optimization targeted at the maximization of convex functions over convex sets, such as cutting plane or branch-and-bound methods~\cite[Ch.~9-11]{Benson_Concave}. 

\section{Hard Pairwise Clusters are Not Always Optimal}\label{sec:pairwisecluster}
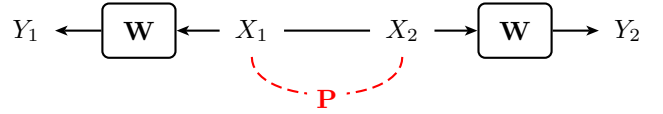
\begin{figure}[t]
 \centering
    \begin{pspicture}[showgrid=false](-4,0.25)(4,1.5)
      \psset{style=RoundCorners}
      \pssignal(-4,1){y1}{$Y_1$}
      \pssignal(-1,1){x1}{$X_1$}
      \pssignal(1,1){x2}{$X_2$}
      \pssignal(4,1){y2}{$Y_2$}
      \ncline{x1}{x2}
      \psfblock[framesize=1 0.75](2.5,1){w2}{$\Wvec$}
      \psfblock[framesize=1 0.75](-2.5,1){w1}{$\Wvec$}
      \nclist[style=Arrow]{ncline}{x2,w2,y2} 
      \nclist[style=Arrow]{ncline}{x1,w1,y1} 
      \ncarc[style=Dash,arcangleA=-90,arcangleB=-90,linecolor=red]{x1}{x2}
      \rput*(0,0.1){\textcolor{red}{$\Pvec$}}
    \end{pspicture}
\caption{Pairwise clustering to maximize $\mutinf{Y_1;Y_2}$ requires finding \emph{a single} matrix $\Wvec$ for both RVs. As Example~\ref{ex:stochastic} shows, deterministic clusters are not always optimal.}
\label{fig:pairwiseclusters}
\end{figure}
 
Suppose now that the data $\mathcal{D}$ consists only of a single set of $N$ points that shall be clustered. Converting this set to a probability matrix $\Pvec$ as in, e.g., Example~\ref{exp:pairwise}, yields two RVs $X_1$ and $X_2$ with the same alphabet. In information-theoretic pairwise clustering, depicted in Fig.~\ref{fig:pairwiseclusters}, one aims at finding a single mapping $\Wvec$ that maps both $X_1$ to $Y_1$ and $X_2$ to $Y_2$, such that the mutual information between the clustered RVs $Y_1$ and $Y_2$ is maximized.

Note that, aside from the clustering problem of Example~\ref{exp:pairwise}, the cost function for pairwise clustering is also used for the aggregation of Markov chains, where $X_1$ and $X_2$ indicate two consecutive samples of the stochastic process. Specifically,~\cite{Goldberger_MarkovAggregation,Meyn_MarkovAggregation} investigate the problem of clustering the alphabet of a Markov chain such that the mutual information between two consecutive state clusters $Y_1$ and $Y_2$ is maximized.

\begin{figure}[t]
 \centering
 \includegraphics[width=0.5\textwidth]{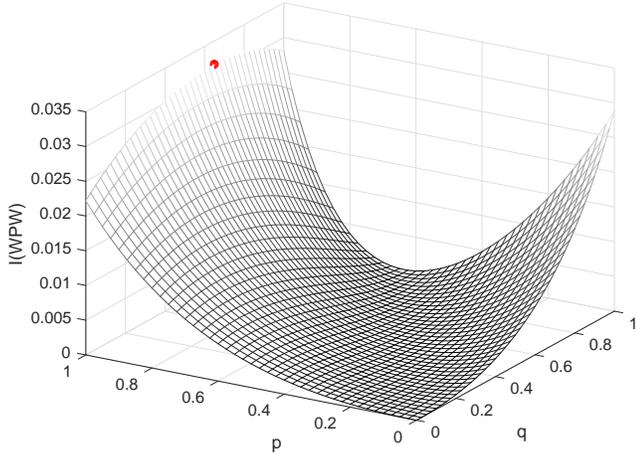}
 \caption{Mutual information $\mutinf{\overline\Wvec\Pvec\Wvec}$ for Example~\ref{ex:stochastic} as a function of $p$ and $q$ for fixed $r=1$. The maximum value (red dot) is achieved for the stochastic matrix $\Wvec$ with $q=0.65$ (see text), while the deterministic matrices are strictly worse.}
 \label{fig:curves}
\end{figure}

Let us now formulate the pairwise clustering problem mathematically. In pairwise clustering we wish to find a maximizer $\Wvec^\bullet$ of
\begin{equation}\label{eq:problem_pairwise}
 \Wvec^\bullet =\argmax_{\Wvec\in\mathcal{M}_{N\times M}} \mutinf{\overline\Wvec\Pvec\Wvec}.
\end{equation}
Although pairwise clustering can thus be seen as restricted co-clustering, cf.~\eqref{eq:problem_coclustering}, our analysis from Theorem~\ref{thm:coclustering} does not carry over to this case. In fact, the following example shows that deterministic pairwise clusters are not sufficient: 

\begin{example}\label{ex:stochastic}
Let $N=3$ and
\begin{equation}
 \Pvec=\left[ \begin{array}{ccc}
                0.1 & 0.1 & 0.175\\
                0.1 & 0.15 & 0.075\\
                0.175 & 0.075& 0.05
               \end{array}\right].
\end{equation}
We wish to cluster $X_1$ and $X_2$ pairwise to $M=2$ clusters, hence we are looking for $\Wvec\in\mathcal{M}_{3\times 2}$ that maximizes the mutual information between $Y_1$ and $Y_2$. We parametrized $\Wvec$ as
\begin{equation}
 \Wvec=\left[ \begin{array}{ccc}
                p & 1-p \\
		q & 1-q\\
		r & 1-r
               \end{array}\right]
\end{equation}
and, in simulations, swept all three parameters $p$, $q$, and $r$ between 0 and 1 in steps of 0.025. The parameters maximizing the mutual information were found to be $p=1$, $q=0.65$, and $r=0$, achieving a mutual information of 0.0316. In comparison, the mutual information obtained by the three (nontrivial) deterministic pairwise clusterings evaluated to 0.0281 for $p=q=1-r=1$, 0.0288 for $p=1-q=r=1$, and 0.0222 for $p=1-q=1-r=1$. These values can be seen as the corner points in Fig.~\ref{fig:curves}, together with the trivial result of a vanishing mutual information for $p=q=r=1$.
\end{example}

\section{When are Optimal Pairwise Clusterings Hard?}\label{sec:whenhard}
In the previous section we showed that, in general, the matrix $\Wvec$ optimal for information-theoretic pairwise clustering is not deterministic. In this section, we present scenarios for which we can either show or for which we have experimental evidence that hard pairwise clusters maximize mutual information. 

\subsection{Diagonal Probability Matrix}\label{subsec:diag}
Suppose that $\Pvec=\diag{\muvec}$, i.e., that $X_1= X_2$. We have by the properties of mutual information,
\begin{equation}\label{eq:diagonal}
 \mutinf{\overline\Wvec\Pvec\Wvec} = \mutinf{Y_1;Y_2} \le \mutinf{X_1;Y_2} = \mutinf{X_2;Y_2}.
\end{equation}
From Section~\ref{sec:cluster}, we know that the right-hand side is maximized for a deterministic matrix $\Wvec$, which implies $Y_1=Y_2$ and achieves equality in~\eqref{eq:diagonal}. Hence, a deterministic matrix solves the pairwise clustering problem~\eqref{eq:problem_pairwise} for a diagonal $\Pvec$.

Looking at the co-clustering problem for a diagonal $\textbf{P}$, one can show that, for any two deterministic matrices $\Vvec^{(1)}$ and $\Vvec^{(2)}$, we have 
\begin{align}\label{eq:detcopair}
  \mutinf{\overline{\Vvec^{(1)}}\textbf{P}\Vvec^{(2)}} &\leq \mutinf{\overline{\Vvec^{(2)}}\textbf{P}\Vvec^{(2)}} = \ent{Y_2}.
\end{align}
In other words, choosing the same deterministic matrix for both RVs is at least as good as choosing two different deterministic matrices. This implies that
\begin{align}
\max_{\Wvec^{(1)},\Wvec^{(2)}\in\mathcal{M}_{N\times M}} \mutinf{\overline{\Wvec^{(1)}}\Pvec\Wvec^{(2)}} 
&= \max_{\Wvec\in\mathcal{M}_{N\times M}} \mutinf{\overline{\Wvec}\Pvec\Wvec}
\end{align}
i.e., the co-clustering and pairwise clustering problems coincide for a diagonal $\textbf{P}$.

\subsection{``Lifted'' Probability Matrix: The Stochastic Block Model}
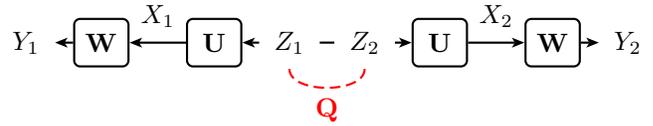
\begin{figure}[t]
 \centering
    \begin{pspicture}[showgrid=false](-4,0.25)(4,1.5)
      \psset{style=RoundCorners}
      \pssignal(-4,1){y1}{$Y_1$}
      \pssignal(-0.5,1){z1}{$Z_1$}
      \pssignal(0.5,1){z2}{$Z_2$}
      \pssignal(4,1){y2}{$Y_2$}
      \ncline{x1}{x2}
      \psfblock[framesize=0.75 0.66](3,1){w2}{$\Wvec$}
      \psfblock[framesize=0.75 0.66](-3,1){w1}{$\Wvec$}
      \psfblock[framesize=0.75 0.66](1.5,1){u2}{$\Uvec$}
      \psfblock[framesize=0.75 0.66](-1.5,1){u1}{$\Uvec$}
      \nclist[style=Arrow]{ncline}[naput]{z2,u2,w2 $X_2$,y2} 
      \nclist[style=Arrow]{ncline}[nbput]{z1,u1,w1 $X_1$,y1} 
      \ncline{z1}{z2}
      \ncarc[style=Dash,arcangleA=-90,arcangleB=-90,linecolor=red]{z1}{z2}
      \rput*(0,0.1){\textcolor{red}{$\Qvec$}}
    \end{pspicture}
\caption{The lifted model assumes that the distribution between $X_1$ and $X_2$ is modeled by a low-rank matrix $\Pvec=\overline\Uvec\Qvec{\Uvec}$. The channel $\Uvec$ is information-preserving, i.e., every state of $Z_1$ is mapped to a different subset of states of $X_1$. By choosing $\Wvec$ as the deterministic matrix that clusters these subsets, one obtains $\mutinf{\overline{\Wvec}\Pvec\Wvec}=\mutinf{\Pvec}=\mutinf{\Qvec}$.}
\label{fig:sbm}
\end{figure}

Consider an arbitrary $M\times M$ probability matrix $\Qvec$ modeling the joint distribution between $Z_1$ and $Z_2$, and let $\Vvec$ be a deterministic $N\times M$ matrix. From this matrix $\Vvec$, we derive a stochastic matrix $\Uvec\in\mathcal{M}_{M\times N}$ with entries
\begin{equation}
 U_{ij} = \frac{\nu_j V_{ji}}{\sum_{k=1}^N \nu_k V_{ki}} \Leftrightarrow \Uvec= \diag{\overline\nuvec\Vvec}^{-1}\overline\Vvec\diag{\nuvec}
\end{equation}
where $\nuvec$ is a length-$N$ vector with positive entries. We now use $\Uvec$ to ``lift'' the joint distribution between $Z_1$ and $Z_2$ to a joint distribution $\Pvec=\overline\Uvec\Qvec{\Uvec}$ between $X_1$ and $X_2$ (see Fig.~\ref{fig:sbm}). One can see that every state of $X_i$ is related to exactly one state of $Z_i$: The channel $\Uvec$ from $Z_i$ to $X_i$ preserves information. It follows that the data processing inequality $\mutinf{\Pvec}\le\mutinf{\Qvec}$ becomes an equality. Since moreover $\Uvec\Vvec=\eye$, choosing $\Wvec=\Vvec$ yields
\begin{equation}
 \mutinf{\overline{\Wvec}\Pvec\Wvec} = \mutinf{\overline{\Vvec}\overline{\Uvec}\Qvec\Uvec\Vvec} = \mutinf{\Qvec}
\end{equation}
i.e., $\Wvec = \Vvec$ achieves the upper bound on mutual information. Hence, a deterministic clustering solves problem~\eqref{eq:problem_pairwise}.

This model appears in the aggregation of Markov chains, e.g.,~\cite{Meyn_MarkovAggregation}, and describes models in which the outgoing probabilities of a state depend only on the state cluster. Specifically, suppose that $\gammavec=\Qvec\onevec=\overline{\overline\onevec\Qvec}$. Then, $\tilde\Qvec=\diag{\gammavec}^{-1}\Qvec$ is the transition probability matrix of a Markov chain with alphabet size $M$. If we choose $\nuvec$ such that $\overline\nuvec\Vvec=\overline\gammavec$, then
\begin{equation}
 \tilde\Pvec=\diag{\nuvec}^{-1}\Pvec = \Vvec\tilde\Qvec\Uvec
\end{equation}
is a transition probability matrix of a Markov chain with alphabet size $N$, that has equal rows for all states belonging to the same cluster. This model is strongly related to the phenomenon of \emph{lumpability}, the scenario in which a deterministic function of a Markov chain has the Markov property.

As a second instance for this model, consider a symmetric $\Qvec$ and $\nuvec=\onevec$. One thus obtains the stochastic block model, a popular model for random graphs. It generalizes the Erd\H{o}s-R\'enyi model by creating communities, determined by $\Vvec$, within which and between which edges exist with probabilities $Q_{ii}$ and $Q_{ij}$, respectively.

\subsection{Boltzmann Probability Matrix: Pairwise Distances}
Suppose you have $N$ data points and a symmetric $N\times N$ matrix $\Dvec$ that collects pairwise distances between these data points, cf.~Example~\ref{exp:pairwise}. Writing $\Pvec\propto e^{-\sigma \Dvec}$, where $\sigma$ is an appropriate scaling factor, yields a symmetric probability matrix with entries reminiscent of the Boltzmann, or Gibbs, distribution. This approach is popular in random-walk based clustering, where the data set is characterized only via pairwise distances or (dis-)similarities~\cite{Tishby_MarkovRelax,Alush_PairwiseClustering}.

Numerical evidence suggests that, for these $\Pvec$, hard pairwise clusters are optimal. More specifically, it appears that relaxing the pairwise clustering problem to a co-clustering problem yields two identical deterministic matrices, i.e., $\Wvec^{(1)\bullet}=\Wvec^{(2)\bullet}$.

\subsection{Gramian Probability Matrix: The Cosine Similarity Model}
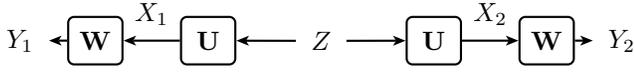
\begin{figure}[t]
 \centering
    \begin{pspicture}[showgrid=false](-4,0.5)(4,1.5)
      \psset{style=RoundCorners}
      \pssignal(-4,1){y1}{$Y_1$}
      \pssignal(0,1){z}{$Z$}
      \pssignal(4,1){y2}{$Y_2$}
      \ncline{x1}{x2}
      \psfblock[framesize=0.75 0.66](3,1){w2}{$\Wvec$}
      \psfblock[framesize=0.75 0.66](-3,1){w1}{$\Wvec$}
      \psfblock[framesize=0.75 0.66](1.5,1){u2}{$\Uvec$}
      \psfblock[framesize=0.75 0.66](-1.5,1){u1}{$\Uvec$}
      \nclist[style=Arrow]{ncline}[naput]{z,u2,w2 $X_2$,y2} 
      \nclist[style=Arrow]{ncline}[nbput]{z,u1,w1 $X_1$,y1} 
    \end{pspicture}
\caption{The Gramian model assumes that $\Pvec\propto\Amat\overline\Amat$, where the rows of $\Amat$ are positive length-$L$ feature vectors. As we argue in the text, $X_1$ and $X_2$ are independent observations of a uniform RV $Z$ via the channel $\Uvec$. In contrary to the lifted model, $\Uvec$ is not information-preserving.}
\label{fig:cosine}
\end{figure}

Consider an $N\times L$ matrix $\Amat$ with all entries being positive. The $N\times N$ Gram matrix $\Amat\overline\Amat$ is symmetric, has positive entries, and has rank at most $\min\{N,L\}$. Normalizing the columns yields a matrix $\Uvec=\diag{\overline\onevec\Amat}^{-1}\overline\Amat\in\mathcal{M}_{L\times N}$, which allows us to write 
\begin{equation}\label{eq:cosinesimilarity}
 \Pvec = \overline\Uvec \Lambdavec\Uvec
\end{equation}
where $\Lambdavec=\diag{\overline\onevec\Amat}^2/\overline\onevec\Amat\overline\Amat\onevec$. Note that Section~\ref{subsec:diag} is a special case for $L = N$ and $\Uvec = \eye$. 

This model is known as the \emph{cosine similarity model}, where $\Amat$ collects length-$L$ feature vectors for each of the $N$ data points. The entries of the Gram matrix $\Amat\overline\Amat$, and hence of $\Pvec$, describe the cosine of the angle between these length-$L$ feature vectors. If the entries of $\Amat$ are all positive, then we can guarantee that $\Pvec$ is a probability matrix.

Preliminary numerical simulations suggest that also for this model, hard pairwise clusters are optimal. Referring to~\eqref{eq:cosinesimilarity}, this model assumes that a single, $L$-valued RV $Z$ is observed twice via a channel $\Uvec$ (see Fig.~\ref{fig:cosine}). Decomposing $\Uvec$ using Lemma~\ref{lem:convcomb} we have
\begin{align}
 \Uvec = \sum_{i} \phi_i \Rvec^{(i)}
\end{align}
where $\Rvec^{(i)}$ are deterministic matrices. As in the proof of Theorem~\ref{thm:coclustering} we can write
\begin{align}
   &\mutinf{\overline\Wvec\Pvec\Wvec}\notag\\
   &\leq  \sum_{i}\sum_{j} \lambda_i \lambda_j \mutinf{\overline{\Vvec^{(i)}}\overline\Uvec\Lambdavec\Uvec\Vvec^{(j)}} \label{eq:firstineq}\\
   &\leq \sum_{i}\sum_{j} \sum_{k}\sum_{l}\lambda_i \lambda_j\phi_k\phi_l \mutinf{\overline{\Vvec^{(i)}}\overline{\Rvec^{(k)}}\Lambdavec\Rvec^{(l)}\Vvec^{(j)}}. \label{eq:secondineq}
\end{align}
Since the product of two deterministic matrices is a deterministic matrix, we obtain an upper bound of~\eqref{eq:secondineq} by
\begin{equation}
  \mutinf{\overline\Wvec\Pvec\Wvec} 
  \leq \max\limits_{\Vvec^{(1)},\Vvec^{(2)}} \mutinf{\overline\Vvec^{(1)}\Lambdavec\Vvec^{(2)}} \overset{(a)}{=} \max\limits_{\Vvec} \mutinf{\overline\Vvec\Lambdavec\Vvec}
\end{equation}
where the maximum is over all $N\times M$ deterministic matrices and where $(a)$ follows from the discussion in Section~\ref{subsec:diag}. In order to show that the pairwise clustering for a $\Pvec$ of the form~\eqref{eq:cosinesimilarity} is deterministic, we need to apply the analysis from Section~\ref{subsec:diag} to the bound~\eqref{eq:firstineq}. A sufficient condition for hard pairwise clusters to be optimal is, if for any two deterministic matrices $\Vvec^{(1)}$ and $\Vvec^{(2)}$,
\begin{multline}
  \mutinf{\overline{\Wvec^{(1)}}\Lambdavec\Wvec^{(2)}} \\
  \leq \max\left\{\mutinf{\overline{\Wvec^{(2)}}\Lambdavec\Wvec^{(2)}}, \mutinf{\overline{\Wvec^{(1)}}\Lambdavec\Wvec^{(1)}}\right\}.
\end{multline}
where $\Wvec^{(i)}=\Uvec\Vvec^{(i)}$.
Unfortunately, we found counterexamples for this statement. 

At the time of submission, a proof that hard pairwise clusters are optimal has still eluded us. Note, however, that in this setting (cf.~Fig.~\ref{fig:cosine})
\begin{equation}
  \mutinf{Y_1;Y_2}
  =\mutinf{Z;Y_1}+\mutinf{Z;Y_2}-\mutinf{Z;Y_1,Y_2}.
\end{equation}
We believe that this result is a good starting point for finding a proof. Future work shall investigate this issue.

\section*{Acknowledgments}
 The authors thank Roy Timo, Technical University of Munich, and Jalal Etesami, University of Illinois Urbana-Champaign, for fruitful discussions.
 This work was supported by the German Ministry of Education and Research in the framework of an Alexander von Humboldt Professorship. The work of Bernhard C. Geiger was funded by the Erwin Schr\"odinger Fellowship J 3765 of the Austrian Science Fund.

\bibliographystyle{IEEEtran} 
\bibliography{IEEEabrv,references}

%

\end{document}